\documentclass[12pt]{article}
\usepackage{subeqn}
\usepackage{epsfig}
\newcommand{\be}{\begin{equation}}
\newcommand{\ee}{\end{equation}}

\newcommand{\vl}{\vskip 0.2in}
\newcommand{\no}{\noindent}
\newcommand{\ce}{\begin{center}}
\newcommand{\nc}{\end{center}}

\makeatletter
\@addtoreset{equation}{section}
\makeatother

\begin{document}

%\rightline{UCLA/03/TEP/--}
%\rightline{October 1999}
%\vskip1.0cm

\begin{center}
{\bf $q$-ELECTROWEAK, $q$-GRAVITY, AND KNOTTED SOLITONS}
\end{center}
\vskip.5cm

\ce{\it R. J. Finkelstein}
\nc
\vskip.3cm
\ce{Department of Physics and Astronomy}
\nc
\ce{University of California, Los Angeles, CA 90095-1547}
\nc
\vskip1.0cm

\no
{\bf Abstract.}   If the Lie group of a non-Abelian theory is replaced by
the corresponding $q$-group, one is led to replace the Lie algebra by two
dual algebras.  The first of these lies close to the Lie algebra that it is
replacing while the second introduces new degrees of freedom.  We interpret
the theory based on the first algebra as a modification of standard field 
theory while we propose that the new degrees of freedom introduced by the
second algebra describe solitonic rather than point particle sources.  We
have earlier found that the modified $q$-electroweak theory differs
very little from the standard theory.  Here we find a similar result
for $q$-gravity.  Both of the modified theories are incomplete, however, and
must be completed by the solitonic sector. 
We propose that the solitonic sector of both 
$q$-electroweak and
$q$-gravity have the symmetry of knots associated with $SU_q(2)$.  Since the
Lorentz group is here deformed, there is no longer the standard classification 
of particles described by mass and spin.  There is instead a classification of
irreducible structures determined by $SU_q(2)$.

\newpage

\no
\section{Introduction.}

\vskip.3cm

Since the $q$- and Lie algebras are closely related, it has been natural to
study the $q$-theories obtained by replacing the Lie algebras in our current
theories and in particular in our description of elementary particles.  In 
doing this a distinction should be made between the Lorentz algebra and the
algebras of the standard model since the latter are phenomenological and less
solidly based than the former.  For example, a $q$-electroweak theory may
be obtained by replacing $SU(2)$ electroweak by $SU_q(2)$ electroweak and at
the same time retaining the Lorentz group.$^1$  In general in going from
$SU(N)$ to $SU_q(N)$ the original algebra gets replaced by two algebras, the
first lying close to the Lie algebra that is being replaced.  The second
algebra and its attached state space introduces new degrees of freedom that can
naturally be associated with non-locality.

We are here using the language of Lie groups rather than Hopf algebras since
we want to emphasize a correspondence limit with standard theory in which
``internal algebra" describes a deformation of the usual Lie group and ``external algebra"
describes a deformation of the usual Lie algebra.  Since the standard Lie group may be obtained by integrating its Lie algebra, all degrees of freedom of the
standard theory are already exposed in the Lie algebra.  That is not true in
the $q$-theory, and is the reason for discussing both algebras here.  

The part of the program associated
with the external algebra has been carried out for $q$-electroweak 
and leads to a
modified Weinberg-Salam theory that is not reducible to the standard
Weinberg-Salam theory and therefore has slightly different experimental
consequences.  This so modified Weinberg-Salam
theory is not renormalizable and needs to be completed
by taking the internal algebra into account.

Here we also discuss the more speculative $q$-Lorentz or $q$-gravity theories.
Both $q$-\break electroweak and $q$-gravity are based on the algebras defined by
\be
T^t \epsilon T = \epsilon 
\ee
\no where
\be
\epsilon = \left(\matrix{0 & q^{-1/2} \cr
-q^{1/2} & 0 \cr} \right) 
\ee
\no For $q$-electroweak $T\epsilon SU_q(2)$ and for $q$-gravity
$T\epsilon SL_q(2)$.

In our view the really interesting feature of the $q$-theories is the appearance of the two
algebras.  It is natural to regard the external algebra, differing little from
the parent Lie algebra, as underlying a modified
standard field theory with point particle sources.  Then the internal algebra
may be interpreted as underlying a
dual description of the same field but with solitonic sources.
It is natural to expect solitons here, since
gauge theories are non-linear theories with attractive self-interactions.
There are numerous examples of solitons in non-linear theories, including
spinor solitons, `t Hooft-Polyakov solitons, and Nambu strings and
other stringlike structures formed by attractive self-interactions.  In our
picture one would expect the external theory to represent a perturbative
description of the full theory, and the internal theory to provide a
non-perturbative description of the full theory as well as a
classification of the solitonic sources.  In the simplest model one may
assume that the two algebras implement the same Lagrangian.

Sections 2, 3, and 4 summarize familiar facts about the spin representation of
the $q$-Lorentz group, $q$-spinors, and $\sigma_q$ matrices.  Section 5
describes the higher dimensional representations of $SU_q(2)$.  In
Section 6 the curvature of standard Euclidean gravity is expressed in terms of the spinor connection of $SU(2)\times SU(2)$ and in Section 7 the relation between
external $q$-gravity and standard gravity is examined.  It is shown there
that the external $q$-gravity is very close to the standard gravity theory
just as external $q$-electroweak is close to the standard electroweak.
Both $q$-electroweak and $q$-gravity are therefore approximately correct
physical theories; but since neither is tree unitary, there must be some
missing physics.  In the present situation it is natural to try to identify
the missing physics with the internal theory.  Section 10 conjectures that the
internal theory may describe closed and knotted flux tubes that play the
role of solitonic sources.

\vskip.5cm

\no
\section{The $q$-Lorentz Group.} 
\vskip.3cm

The formalism of $q$-gravity resembles $q$-electroweak theory in that the
affine connection $\Gamma_\mu$ of $q$-gravity lies in $SL_q(2)$ while the
corresponding vector potential $A_\mu$ of $q$-electroweak lies in $SU_q(2)$,
a subgroup of $SL_q(2)$.

Let us recall the $q=1$ limit of $q$-Lorentz by introducing
\be
X = x_k\sigma_k \qquad k = 0,1,2,3 
\ee
\no where the $x_k$ are real and $\sigma_k = (1,\vec\sigma)$ so that
\be
X = \left(\matrix{t+z & x-iy \cr
x+iy & t-z \cr} \right) 
\ee
\no Then
\be
X^+ = X 
\ee
\no and
\be
{\rm det}~X = t^2-x^2-y^2-z^2 
\ee
\no Now introduce the 2-dimensional representation of the Lorentz
transformation $(L)$ by setting
\be
L = \left(\matrix{a & b \cr
c & d \cr} \right) 
\ee
\no where the matrix elements of $L$ are complex and restricted by
\be
{\rm det}~L = 1 
\ee
\no By (2.6) the number of independent real parameters of $L$ is reduced
to six, the number needed to characterize a Lorentz transformation.

Now transform $X$ by
\be
X^\prime = L^+XL 
\ee
\no where $L^+$ is $L$ adjoint.  Then

\be (X^\prime)^+ = X^\prime
\ee
\be 
{\mbox det}~X^\prime = {\mbox det}~X 
\ee
\be 
(x_o^2-\vec x^2)^\prime = (x_o^2-\vec x^2) 
\ee 

\no Hence the 6 independent parameters of $L$ may be identified with the
parameters of a Lorentz transformation.

The unimodular restriction on $L$ may be expressed as follows:
\be
\epsilon_{ij}L_{im}L_{jn} = \epsilon_{mn}{\rm det}~L = \epsilon_{mn}
\ee
\no or
\be
L^t\epsilon L = \epsilon 
\ee
\no where
\be
\epsilon_{mn} = \left(\matrix{0 & 1 \cr -1 & 0 \cr} \right) 
\ee
\no One may pass to the $q$-theory by replacing $\epsilon_{mn}$ by
\be
\epsilon_{mn}(q) = \left(\matrix{0 & q_1^{1/2} \cr
-q^{1/2} & 0 \cr} \right) \qquad q_1 = q^{-1} 
\ee
\no and requiring the analog of (2.12)
\be
L_q^t\epsilon(q)L_q = L_q\epsilon_q L_q^t = \epsilon_q 
\ee
\no The matrix elements of $L_q$ must now satisfy the following algebra:

\be
\begin{array}{llll} 
  ab = qba & ac = qca & bc = cb & ad-qbc = 1 \\
bd = qdb & cd = qdc  & &  da-q_1bc = 1  
\end{array}
\ee

The $q$-determinant is
\be
ad-qbc 
\ee
\no the natural generalization of (2.11) and (2.14).

\vl

\section{$q$-Spinors.$^2$}
\vskip.3cm

We shall next drop the subscripts on $L_q$ and $\epsilon_q$ and understand the symbols $L$ and $\epsilon$ to represent the $q$-deformed matrices.

Let $\psi^A$ be a contravariant 2-rowed basis and $\chi^{\dot A}$ a
contravariant basis for the conjugate representation:
\be
\psi^{A^\prime} = L^A_{~B}\psi^B
\ee
\be 
\chi^{\dot A^\prime} = (L^\star)^{\dot A}_{~\dot B} \chi^{\dot B} 
\ee
\no Associated with $\psi^A$ and $\chi^{\dot A}$ are corresponding covariant
spinors
\be
\psi_A^\prime = \psi_B(L^{-1})^B_{~A}
\ee
\be 
\chi_{\dot A}^\prime = \chi_{\dot B}\bigl((L^\star)^{-1}\bigr)^{\dot B}_{~\dot A}
\ee
\no By (2.15) $\epsilon_{AB}$ is an invariant second rank tensor since
\be
\epsilon_{AB} = L_A^{~C}L_B^{~D}\epsilon_{CD} 
\ee
\no Then $\psi^A\epsilon_{AB}\chi^B$ is an invariant form:
\be
\psi^{A^\prime}\epsilon_{AB}\chi^{B^\prime} = (L^A_{~C}\psi^C)
\epsilon_{AB}(L^B_{~D}\chi^D)
\ee
\[
= \psi^C\bigl((L^t)_C^{~A}\epsilon_{AB}L^B_{~D}\bigr)\chi^D \] 
\be
= \psi^C\epsilon_{CD}\chi^D 
\ee
\no while
\begin{eqnarray*}
(\psi^A\chi^{\dot X})^\prime &=& (L^A_{~B}\psi^B)(L^\star)^{\dot X}_{~\dot Y}
\chi^{\dot Y} \\ 
&=& L^A_{~B}(\psi^B\chi^{\dot Y})L^{+~\dot X}_{~\dot Y} 
\end{eqnarray*}
\no Then
\begin{subequations}
\[
M^{A\dot X^\prime} = L^A_{~B}M^{B\dot Y}(L^+)_{\dot Y}^{~\dot X}
\]
\no or
\be
M^\prime = L~M~L^+
\ee
\no where
\be 
M^{A\dot X} = \psi^A\chi^{\dot X} 
\ee
\end{subequations}

\no Now the remarks following (2.7) do not hold for (3.8a) since the matrix
elements of $L$ no longer commute.  In particular, det $M$ and $x_o^2-\vec x^2$ are no longer invariant.

The general $q$-spin tensor may again be defined as

\be
u(n,m) = \psi_{k_1\ldots k_n}\chi_{\dot\ell_1\ldots\dot\ell_m}
\ee
\no that transforms like the product of $n$ $q$-spinors and $m$ complex 
conjugate $q$-spinors.  Then $u(n,m)$ is the basis of an irreducible
representation of the $q$-Lorentz group.

We define a contravariant $\epsilon$ symbol by
\be
\epsilon^{AB}(q) = \epsilon_{AB}(q^{-1}) 
\ee
\no and $\epsilon$ may be used to raise or lower indices to obtain
contra- or covariant spin tensors.
\vskip.5cm

\section{The $\sigma_q$ Matrices.$^2$} 
\vskip.3cm

Let
\be
(\sigma_q^m)_{B\dot Y} = (1,\vec\sigma) 
\ee
\no be the usual Pauli matrices.  Introduce the matrices contravariant to
$(\sigma_q^m)_{B\dot Y}$ with respect to the metric $\epsilon_q$:
\be
(\bar\sigma_q^m)^{\dot XA} = \epsilon_q^{\dot X\dot Y}\epsilon_q^{AB}
(\sigma_q^m)_{B\dot Y} 
\ee
\no Then
\be
(\bar\sigma_q^m)^{\dot XA} = 
\left(\matrix{q & 0 \cr 0 & q^{-1} \cr}\right) \quad
\left(\matrix{0 & -1 \cr -1 & 0 \cr}\right) \quad
\left(\matrix{0 & i \cr -i & 0 \cr} \right) \quad
\left(\matrix{-q & 0 \cr 0 & q^{-1} \cr} \right) 
\ee
\no by (3.10).  These matrices satisfy the following relations$^3$
\be
(\bar\sigma_q^m)^{\dot XA}(\sigma_q^n)_{A\dot X} =
2\eta^{mn}
\ee
\be 
(\sigma_q^n)_{A\dot X}(\bar\sigma_{qn})^{\dot YB} =
2\delta^{\dot Y}_{\dot X}\delta^B_A 
\ee 

\no where
\begin{subequations}
\be
\eta^{nm} = \left(\matrix{{1\over 2}(q+q^{-1}) & 0 & 0 & {1\over 2}(q-q^{-1})
\cr 
0 & -1 & 0 & 0 \cr 0 & 0 & -1 & 0 \cr
-{1\over 2}(q-q^{-1}) & 0 & 0 & -{1\over 2}(q+q^{-1}) \cr} \right)
\ee 
\no
and
\be
{\rm det}~\eta = -\cosh^2\theta + \sinh^2\theta = -1 
\ee
\no
where
\be
q = e^\theta
\ee
\end{subequations} 

\no The equation of the light cone is then

\[
\eta_{nm}x^nx^m = \eta_{00}c^2t^2 + \eta_{33}z^2-x^2-y^2 = 0 \]
\be 
= (\cosh\theta)(c^2t^2-z^2)-x^2-y^2 = 0 
\ee  

\no or
\begin{subequations}
\be
c^2\tau^2-\zeta^2-x^2-y^2 = 0 
\ee
\no after the rescaling where
\[
\zeta = (\cosh\theta)^{1/2}z \]
\be
\tau = (\cosh\theta)^{1/2} t 
\ee
\end{subequations}

\no The light cone is then rescaled in the $q$-theory, but it is not
invariant under the deformed Lorentz transformation.

By (3.7) the spinor bilinear $\psi^t\epsilon\psi$ is invariant under $q$-Lorentz
transformations satisfying (2.15).

The basic invariant is therefore no longer the interval or the light cone
but is instead the spinor cone or the associated $q$-commutor
\begin{subequations}
\be
\psi^t\epsilon\psi = 0
\ee 
\no or
\be
(\psi^1,\psi^2)_q \equiv \psi^1\psi^2-q\psi^2\psi^1 = 0 
\ee
\end{subequations}

Since Einstein-Minkowski spacetime is based on the metrical light cone and since
that is not invariant under $q$-Lorentz transformations, it must be replaced
here by a $q$-spacetime that is based on the spinor cone that is invariant
under these transformations.  Hence if $q\not= 1$, the light cone and an
associated Einstein-Minkowski space are replaced by a non-commuting algebra
and associated non-commuting analogue of group space, as discussed by many
authors.

At the same time the elementary particle states, labeled by mass and spin,
and defined by the Lorentz and Poincar\'e algebras, disappear, and one is left
with only structures and states defined by the $SL_q(2)$ algebra.
\vskip.5cm

\section{The Irreducible Representation of $SU_q(2)$.} 
\vskip.3cm

Instead of deforming the spin representation of the Lorentz group one may
deform its $SO(4)$ representation.  Since $SO(4) = SU(2) \times SU(2)$ we
may therefore make use of $SU_q(2)$.  The irreducible representations of
$SU_q(2)$ are as follows:$^{3,4}$
\[
D^j_{mm^\prime}(a,\bar a,b,\bar b) = \Delta^j_{mm^\prime}
\sum_{s,t}\biggl\langle\matrix{n_+ \cr s\cr}\biggr\rangle_1
\biggl\langle\matrix{n_- \cr t\cr} \biggr\rangle_1 q^{t(n_++1-s)}
(-1)^t\delta(s+t,n_+^\prime) \] 
\be
\times a^sb^{n_+-s}\bar b^t\bar a^{n_--t} 
\ee
\no where

\[
n_\pm = j\pm m \]
\[ 
n^\prime_\pm = j\pm m^\prime 
\]
\[
\biggl\langle\matrix{n \cr s\cr} \biggr\rangle_1 =
{\langle n\rangle_1!\over\langle s\rangle_1!
\langle n-s\rangle_1!} \quad {\mbox and} \quad
\langle n\rangle_1 = {q_1^{2n}-1\over q_1^2-1} 
\]
\[
\Delta^j_{mm^\prime}= \biggl[{\langle n_+^\prime\rangle_1!~
\langle n^\prime_-\rangle_1!\over\langle n_+\rangle_1!~\langle n_-\rangle_1!}
\biggr]^{1/2} \qquad q_1 = q^{-1}
\]

\no and the arguments of (5.1) satisfy the following relations
\be
\begin{array}{lll}
ab = qba &  a\bar a+b\bar b = 1 & b\bar b = \bar bb \\
a\bar b = q\bar ba & \bar aa + q_1^2\bar bb = 1 & \\ 
\end{array}
\ee
\no if $q$ is real.
These relations are obtained from the corresponding relations for
$SL_q(2)$ by setting$^4$
\[
c = -q_1\bar b \]
\be
a = \bar d
\ee

\no in (2.16).  Now set
\begin{subequations}
\be
D^{1/2}(a,\bar a,b,\bar b) = e^{B\sigma_+}e^{\lambda\theta\sigma_3}
e^{C\sigma_-}
\ee 
\no and expand to terms linear in $(B,C,\theta)$.  Here
\be
q = e^\lambda \quad b = Bq^{-\theta} \quad
q_1\bar b = -q^{-\theta} C 
\ee
\end{subequations}

\no Then

\be
D^j_{mm^\prime}(B,C,\theta) = D^j_{mm^\prime}(0,0,0) + B(J_B^j)_{mm^\prime}
+ C(J_C^j)_{mm^\prime} + 2\lambda\theta(J_\theta^j)_{mm^\prime} + \ldots
\ee
\no The non-vanishing matrix coefficients $(J_\theta^j)_{mm^\prime},
(J_C^j)_{mm^\prime}$, and $(J_\theta^j)_{mm^\prime}$ are by (5.1)

\[
(m-1|J_B^j|m) = [\langle j+m\rangle_{q_1^2}
\langle j-m+1\rangle_{q_1^2}]^{1/2} \]
\be
(m+1|J_C^j|m) = [\langle j-m\rangle_{q_1^2}
\langle j+m+1\rangle_{q_1^2}]^{1/2} \;\;\;\; q_1 = q^{-1} 
\ee
\[
(m|J_\theta^j|m) = m 
\]
\no Then $(B,C,\theta)$ and $(J_B,J_C,J_\theta)$ are generators of two dual
algebras satisfying the following commutation rules

\be
\begin{array}{lll}
(J_B,J_\theta) = -J_B & (J_C,J_\theta) = J_C &
(J_B,J_C) = q_1^{2J-1}[2J_\theta] 
\end{array}
\ee
\be
\begin{array}{lll}
(B,C) = 0 & (\theta,B) = B & 
(\theta,C) = C 
\end{array}
\ee
\no where
\be
[x] = {q^x-q_1^x\over q-q_1} \;\;\;\;\;
\langle x\rangle = {q^x-1\over q-1}
\ee

\no Here the internal algebra is described by (5.2), (5.4) and (5.8)
and the external algebra by (5.7).  These are the two algebras previously
introduced.

The following commutation relations are implied by (5.6)

\no $J = {1\over 2}$ (fundamental)

\be
(J_B,J_\theta) = -J_B \qquad
(J_C,J_\theta) = J_C \qquad
(J_B,J_C) = 2J_\theta 
\ee

\vskip.3cm
\no $J=1$ (adjoint)
\be
(J_B,J_\theta) = -J_B \qquad
(J_C,J_\theta) = J_C \qquad
(J_B,J_C) = \langle 2\rangle_{q_1^2}J_\theta
\ee

\no If $J>1$, the right-hand side of (5.7) is not linear in the generators
and in that case one cannot speak of structure constants or a deformed Lie
algebra.

In the Weinberg-Salam electroweak theory the vector potential lies in the Lie
algebra of $SU(2)$.  In the $q$-electroweak theory it lies in the external
algebra of $SU_q(2)$.  Since only the fundamental and adjoint representations
of the external algebra
are needed, the $q$-electroweak theory differs from standard electroweak
theory only in the adjoint representation, and there only slightly.
We regard this theory, based on the external algebra, as a perturbative
version of the full $q$-electroweak theory that is based on the internal
algebra.

In the corresponding gravitational case, the affine connection must lie
in the Lie algebra 
of $SL(2)$ or in the Euclidean version in $O(4)$, or $SU(2)\times SU(2)$ and in the
$q$-gravitational theory we shall consider only the
$SU_q(2)\times SU_q(2)$ representation.
\vskip.5cm

\section{Euclidean Gravity.}
\vskip.3cm

The $SO(4)$ group may be factored in the well known way, namely:
\be
SO(4) = SU(2)\times SU(2) 
\ee
\no or
\be
e^{i\theta_{\mu\nu}M_{\mu\nu}} = e^{i\theta_{oj}L_j}
e^{i\theta_{ki}J_{ki}} \qquad 
\begin{array}{l}
i,j,k = 1,2,3 \\
\mu\nu = 0,1,2,3
\end{array}
\ee

\no where

\be
(L_i,L_j) = i\epsilon_{ijk}L_k
\ee
\be
(J_i,J_j) = i\epsilon_{ijk}J_k 
\ee
\be
(J_i,L_k) = 0
\ee

\no and
\be
J_k = {1\over 2} \epsilon_{kij}J_{ij}  \qquad \qquad i,j,k = 1,2,3
\ee

\no Here $L_k$ and $J_k$ are both generators of $SU(2)$ and the
$\epsilon_{k\ell s}$ are the structure constants of $SU(2)$.

Let us express the $SU(2)\times SU(2)$ spin connection as follows:
\be
\omega_\mu = \omega_\mu^{oi}s_i + \omega_\mu^{jk}s_{jk} \qquad
i,j,k = 1,2,3
\ee
\no where $s_i$ and $s_{jk}$ are two-dimensional representations
of $L_i$ and $J_{jk}$ respectively.  Then the curvature is
\begin{eqnarray}
R_{\mu\lambda} &=& \partial_\mu\omega_\lambda-\partial_\lambda\omega_\mu
+ (\omega_\mu,\omega_\lambda) \\
&=& R_{\mu\lambda}^{ok}s_k + R^{jk}_{\mu\lambda}s_{jk}
\end{eqnarray}
\no where
\begin{eqnarray}
R^{ok}_{\mu\lambda} &=& \partial_\mu\omega_\lambda^{ok}-\partial_\lambda
\omega_\mu^{ok} +
\epsilon^{ok}_{oi,oj}\omega^{oi}\omega^{oj} \\
R_{\mu\lambda}^{jk} &=& \partial_\mu\omega_\lambda^{jk}-\partial_\lambda
\omega_\mu^{jk} + i~\epsilon^{jk}_{rs,mn}
\omega^{rs}\omega^{mn}
\end{eqnarray}

Here $\epsilon^{ok}_{oi,oj}$ and $\epsilon^{jk}_{rs,mn}$ restate
$\epsilon_{ijk}$ according to (6.3) and (6.4).
Define the antisymmetric matrix $R^{ab}_{\mu\lambda}$ as
follows:
\be
[R^{ab}_{\mu\lambda}] =
\left[ 
\begin{array}{cccc}
0 & R^{01}_{\mu\lambda} & R^{02}_{\mu\lambda} & R^{03}_{\mu\lambda} \\
  & 0 & R^{12}_{\mu\lambda} & R^{23}_{\mu\lambda} \\
  &   & 0 & R^{23}_{\mu\lambda} \\
  &   &   & 0 
\end{array}
\right] \qquad a,b = 0,1,2,3
\ee
\no where
\be
R^{ab}_{\mu\lambda} = -R^{ba}_{\mu\lambda}
\ee
\no Then the action for Euclidean gravity is
\begin{eqnarray}
S &=& \int V^a \wedge V^b \wedge \epsilon_{abcd} R^{cd} \\
&=& \int R \sqrt{-g}~d^4x \qquad {\rm if}~~\det \eta = -1
\end{eqnarray}
\no where $V^a$ and $R^{cd}$ are tetrad and curvature forms.                  

\vskip.5cm

\section{$q$-Gravity.} 
\vskip.3cm

The $q$-gravitational action is
\be
S = \int R\sqrt{-g}~ d^4x 
\ee
\no where the metric $g_{\alpha\beta}(q)$ may be written in terms of the
tetrad $V_\alpha^a$:
\be
g_{\alpha\beta}(q) = V_\alpha^a \eta_{ab}(q)V^b_{~\beta}
\ee
\no and $\eta_{ab}(q)$ is given by (4.6).  Since
\be
{\rm det}~\eta(q) = -1
\ee
\no
we have by (7.2)
\be
{\rm det}~g(q) = -({\rm det}~V)^2
\ee

\no The Riemann tensor is
\be
R_{\mu\lambda}(q) = \partial_\mu\Gamma_\lambda(q)-
\partial_\lambda\Gamma_\mu(q) + [\Gamma_\mu(q),\Gamma_\lambda(q)]
\ee
\no
where
\be
\Gamma^\mu_{\alpha\beta}(q) = {1\over 2}g^{\mu\sigma}(q)
(\partial_\alpha g_{\beta\sigma}(q)+\partial_\beta g_{\alpha\sigma}(q)
-\partial_\sigma g_{\alpha\beta}(q))
\ee

\no By (7.2) the field equations obtained from (7.1) will depend on $q$,
if written in terms of $V^a_{~\alpha}$.  If they are written in terms of the
traditional $g_{\alpha\beta}$ they will be unchanged.  It is then only the
relation between $g_{\alpha\beta}$ and $V_\alpha^a$ that depends on $q$ as
determined by (7.2).

To bring the presentation of $q$-gravity closer to $q$-electroweak, express
the curvature in terms of the spin connection (6.7) rather than the
Christoffel connection (7.6).

To make this transition to the $q$-theory rewrite (6.7), (6.10), and
(6.11) as
\begin{eqnarray}
\omega_\mu(q) &=& \omega_\mu^{oi}(q)s_i(q) +
\omega_\mu^{jk}(q) s_{jk}(q) \\
R_{\mu\lambda}^{ok}(q) &=& \partial_\mu\omega_\lambda^{ok}(q) -
\partial_\lambda\omega_\mu^{ok}(q) + i~\epsilon^{ok}_{oi,oj}(q)
\omega^{oi}(q)\omega^{oj}(q) \\
R^{jk}_{\mu\lambda}(q) &=& \partial_\mu\omega_\lambda^{jk}(q) -
\partial_\lambda\omega_\mu^{jk}(q) + i~\epsilon^{jk}_{rs,mn}(q)
\omega^{rs}(q)\omega^{mn}(q)
\end{eqnarray}
\no But
\be
s_{jk}(q) = s_{jk} \qquad s_k(q) = s_k 
\ee
\no Then
\be
\epsilon_{m\ell s}(q) = \epsilon_{m\ell s} 
\ee  

\no since the fundamental representation is not changed in passing to the
$q$-algebra (see (5.10)).  Therefore $\epsilon^{ok}_{oi,oj}(q)$ and
$\epsilon^{jk}_{rs,mn}(q)$ are also independent of $q$.

Hence this version of $q$-gravity also agrees with Einstein gravity.
Therefore whether one attempts to $q$-deform either the Christoffel
connection $(\Gamma)$ or the spin connection $(\omega)$ the equations
of the free gravitational field are unchanged.

\vskip.5cm

\section{Interacting Fields.}
\vskip.3cm

If one considers the total $q$-field characterized by an action describing
interacting gravitational, electroweak, and spinor fields, then it would
be natural to assume that all groups, including the Lorentz group, are
$q$-deformed. In this action the $q$-deformation will appear in terms describing
the free electroweak field, in interactions of the electroweak with the
gravitational field via the electroweak energy momentum tensor, and in interactions of
both electroweak and gravity with the spinor fields if the spinor interactions
are described by
\be
\psi^t\epsilon_q D_i\psi \qquad i = 1,2 
\ee

\no where $\epsilon_q$ plays the role of the usual charge conjugation matrix
and
\[
D_1 = \gamma^\mu(\partial_\mu+eA_\mu)
\]
\be
D_2 = \gamma^\mu(\partial_\mu+g\omega_\mu)
\ee

\no where $D_1$ and $D_2$ are electroweak and gravitational covariant
derivatives.  Then

\[
(\psi^t\epsilon_qD_i\psi)^\prime = (\psi^tL_q^t)\epsilon_q
(L_qD_iL_q^{-1})L_q\psi
\]
\be
= \psi^t\epsilon_qD_i\psi 
\ee

\no since

\be
L_q^t\epsilon_qL_q = \epsilon_q 
\ee
\be
D_i^\prime = L_qD_iL_q^{-1}
\ee

In the $q$-electroweak case, $\epsilon_q$ is two-dimensional, while in the
$q$-gravity case, it is four-dimensional to agree with $\omega_\mu$.

\vskip.5cm

\section{The Internal Algebra.} 
\vskip.3cm

One sees that $q$-gravity like $q$-electroweak differs little or not at all
from the standard theories in the sector exclusively dependent on the
external algebra.  One may next speculate about the theory based on
the internal algebra.

If one postulates $q$-Lorentz, there is no longer a local Poincar\'e
group permitting the definition of elementary particles in terms of mass and
spin.  Instead the irreducible structures should be defined by 
the irreducible representations of the 
$q$-algebra.
Alternatively the elementary structures may be described as
knotted loops defined by the $q$-algebra.  These knots may be labeled by
their Jones polynomials and these polynomials may be generated directly by
a recipe based on the $q$-algebra.$^5$

The 3-dimensional knots may be characterized by their projections on a
plane, where they appear as four-valent plane graphs with extra structure
at the vertices in the form of the two types of crossings, shown
in figure 1.

\begin{figure}
\epsfxsize=2.5in
\begin{center}
\leavevmode
\epsffile[0 320 249 474]{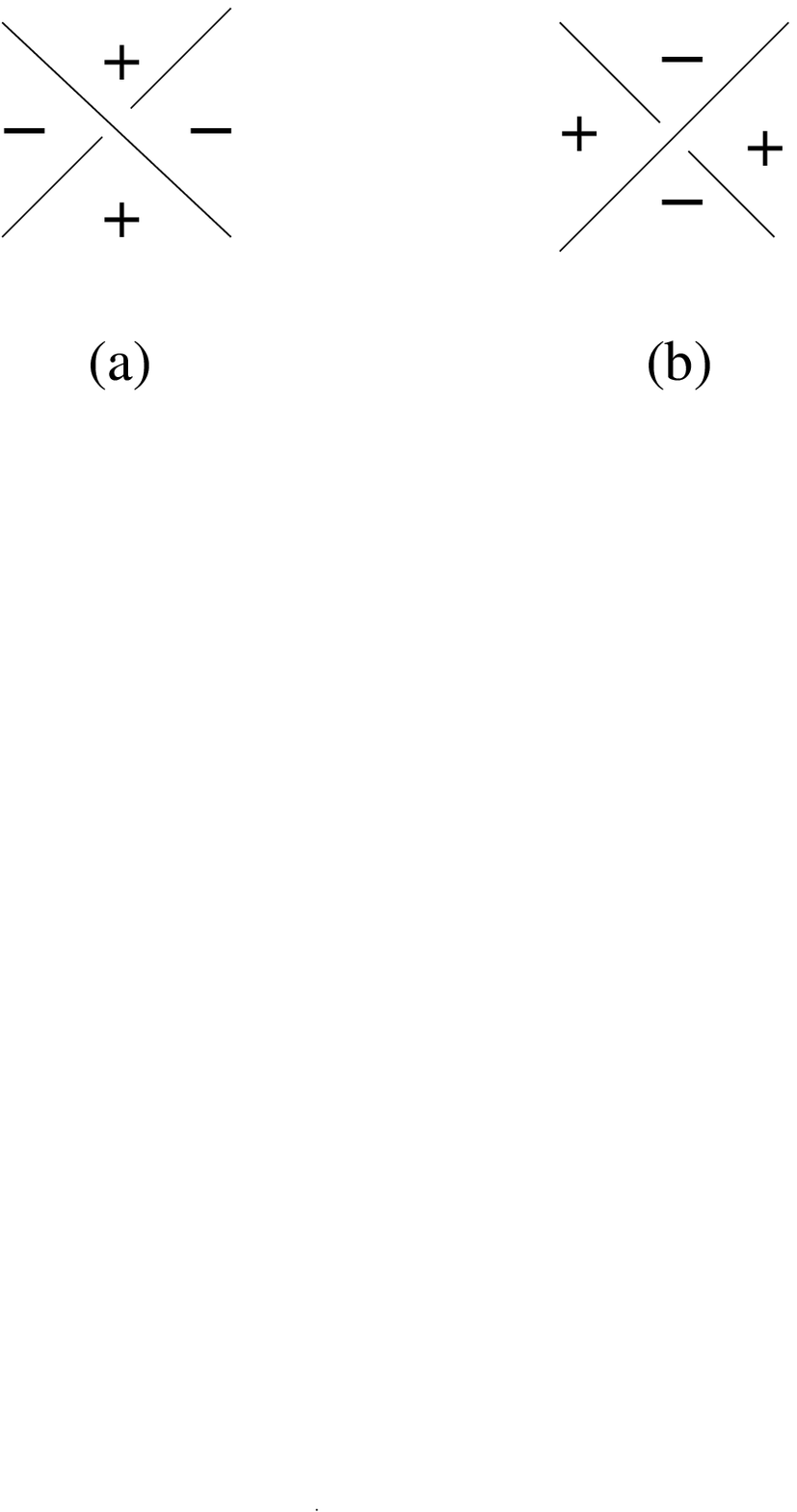}
\end{center}
\caption{}
\end{figure}

The broken arc pair at a crossing indicates the arc that passes underneath
the other arc in space.  If (a) and (b) in Fig. 1 are rotated 
counterclockwise and clockwise respectively, so that the 
overcrossing line lies along the x-axis, then the $(--)$ channel is
composed of the conventional first and third quadrants.

L. Kauffman has shown how to encode a program for generating the Jones
polynomial with the matrix $\epsilon_q$ defined in (1.2).  He
associates a well-defined polynomial $\langle K\rangle$ with an unoriented
link $K$.  This polynomial is defined recursively in Eqs. (9.1), (9.2), and
(9.3):$^5$

\be
\langle K\rangle=\imath \left[q^{-\frac{1}{2}}\langle K_-\rangle -q^{\frac{1}{2}}\langle K_+\rangle \right] 
\ee

\no
where $K$,$K_-$, and $K_+$ are shown in figure 2.

\be
\langle 0~K\rangle = (q+q^{-1})\langle K\rangle 
\ee
\be
\langle 0\rangle = q+q^{-1}
\ee

In the first formula (9.1) brackets like
\be
\biggl\langle \matrix{\ldots \cr X \cr} \biggr\rangle 
\ee
\no refer to graphs with one crossing highlighted.

\begin{figure}
\epsfxsize=4in
\begin{center}
\leavevmode
%\epsffile[0 300 249 474]{fig1.eps}
\epsffile{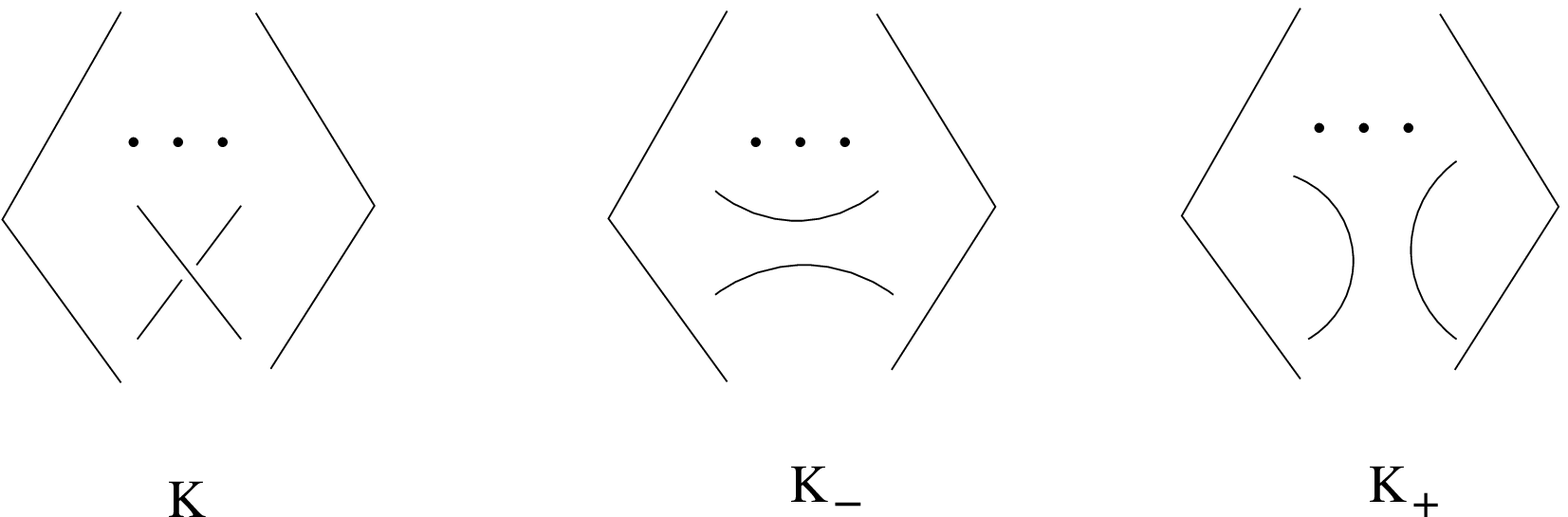}
\end{center}
\caption{}
\end{figure}

Formula (9.1) asserts that the polynomial for a given diagram is obtained
by an additive combination of the polynomials for the diagrams obtained by
splicing away the given crossing in the two possible ways, i.e. one may open up 
either the $q^{-1/2}$ channel or the $q^{1/2}$ channel.  The small diagrams
indicate larger diagrams that differ only as indicated.  Formulas 
(9.2) and (9.3) state that the value of a loop (simple closed curve in the
plane) is $(q+q^{-1})$ and that if the loop occurs (isolated) inside a large
diagram then the value of the polynomial acquires a factor $(q+q^{-1})$ from the
loop.

Repeated applications of these rules to a graph $K$ with multiple crossings
reduce $\langle K\rangle$ to a Laurent polynomial in $q$.

If $K$ is oriented, then one may form the following invariant of ambient
isotopy:
\be
f_k(iq^{-1/2}) = (iq^{-1/2})^{-W(K)}\langle K\rangle
\ee

\no where $W(K)$, the sum of the crossing signs, is the writhe of $K$.
For any oriented link $V_k(t) = f_k(t^{-1/4})$ is the one variable
Jones polynomial.$^5$

Written entirely in terms of $\epsilon_q$ the Kauffman rules (9.1), (9.2),
and (9.3) read as follows:
\begin{eqnarray*}
\langle K\rangle &=& {\rm Tr}~\epsilon_q[\sigma_-\langle K_-\rangle + \sigma_+\langle K_+\rangle] \qquad \sigma_\pm =
\frac{1}{2}(\sigma_1\pm i~\sigma_2) \\
\langle OK\rangle &=& {\rm Tr}~\epsilon_q^t\epsilon_q\langle K\rangle \\
\langle K\rangle &=& {\rm Tr}~\epsilon_q^t\epsilon_q
\end{eqnarray*}

\vskip.5cm

\section{Summary and Remarks.} 
\vskip.3cm

As we have seen in Section 2, $\epsilon_q$ is the basic invariant
of $SL_q(2)$ and hence of the $q$-Lorentz group.  Now we see that it also
encodes the Kauffman bracket.
If the bracket is physically interpreted as a vacuum expectation value,
then $q+q^{-1}$ also acquires a physical meaning as the vacuum expectation value of a loop.

One may establish a correspondence between the irreducible
representations of the $q$-algebra, given by (5.1) and labeled by
$(j,m,m^\prime)$, and the characterization of 
oriented knots by $(N,w,r)$ where $N$
is the number of crossings, $w$ is the writhe, and $r$ is the Whitney degree
or the rotation number of the underlying plane curve.  Then we may 
characterize the knot $(N,w,r)$ by $D^N_{wr}$, also given by (5.1), as
well as by the Jones polynomial.  The Jones polynomial is numerically
valued, while $D^N_{wr}$ is an operator that may be evaluated on the
state space attached to the $q$-algebra.

Here $N$, $w$, and $r$ are all integers. In order to include the
half-integer representations one writes
\be
D^{\frac{N}{2}}_{\frac{w}{2} \frac{r+1}{2}}
\ee
\no
where $N$, $w$, and $r$ are all integers. 

Then one may satisfy the knot constraint:
\be
r-w=\mbox{odd}
\ee

Since the external algebra differs little from the parent Lie algebra 
it may be chosen as
the basis of a modified standard theory that describes point particles.  It
is then natural to associate the additional degrees of freedom of the
internal algebra with the
soliton that replaces the point particle, since there is no other place 
for them in a particle-like description.  Indeed if we exclude
point particles from the theory, the fields must be their own sources.  This
point of view is supported by the existence of solitons in various 
classical non-linear
field theories, including in particular gauge 
theories that exhibit 't Hooft-Polyakov solitons 
or alternatively cohesive flux tubes or strings.  The solitons
usually considered are globular or unknotted, but knotted flux tubes are obvious
possibilities for representing solitonic sources that are localized at some scale 
and we see from the present work that they are natural in $q$-gauge theories.
One may also entertain the possibility that the external theory is a
perturbative version of the internal theory.

One may then conjecture a quantum theory in which the states are states of
knots rather than states of Lorentz particles.  We may describe the knots as
$q$-Lorentz particles characterized by Jones polynomials or by $D^{N/2}_{{\frac{w}{2}}\frac{r+1}{2}}$.

We note that knot states have also emerged from attempts to quantize general\break 
relativity.$^{6,7}$  These results have been summarized as follows:$^6$

In the canonical formulation of Einstein gravity one may take the dynamical
variables to be the spatial components of the Ashtekar connection,
$A_i$, and the corresponding components of the conjugate momentum, $E^i$, 
to be the
densitized triad.  Then in the absence of cosmological and energy-momentum
terms, the constraints are all first class and take the form:

\be
D_iE^i = 0 \qquad \qquad 
\quad ~~\hbox{(gauge invariance)}
\ee
\be
{\rm Tr}~F_{ij}E^i = 0 \qquad \qquad \hbox{(3D-reparametrization)} 
\ee 
\be
{\rm Tr}~F_{ij}E^iE^j = 0 \qquad \quad \hbox{(time reparametrization)}
\ee
\no where $F_{ij}$ is the curvature computed from $A_i$.

The quantum state satisfying all these quantum constraints may be
exhibited as the following integral transform$^7$

\begin{subequations}
\be
\psi(\gamma) = \int {\cal{D}}A~ W(\gamma,A)\psi(A)
\ee

\no where the kernel is the Wilson loop

\be
W(\gamma,A) = {\rm Tr}~{\cal{P}}~e^{\oint_\gamma A}
\ee
\end{subequations}

\no and $\gamma$ is a non-self-intersecting smooth closed curve while $A$
is the Ashtekar connection.  Eq. (10.6) transforms functionals of $A$
into functionals of loops.  Since the diffeomorphism class of a smooth
non-selfintersecting loop is called a knot, the functions of knot classes
satisfy all the constraints of uncoupled quantum general relativity.

From the viewpoint of this paper, however, knotted solitons
should appear in any $q$-gauge theory
with attractive self-interactions.  Here the $\epsilon_q$
symmetry appears as input, but in the quantum gravity work the knot group
appears as output of the quantization.  

Because external $q$-gravity agrees with standard gravity (as described in
Sections 6 and 7), knot solutions to the quantum constraints should appear
in external $q$-gravity as well.  In addition, as shown in Sections 9 and 10,
knots also appear in the internal $q$-theory via $\epsilon_q$, the fundamental
invariant of $q$-Lorentz or $SU_q(2)$.  Knots therefore appear in both the 
internal and external sectors of $q$-gravity.  Since all fields are coupled to the gravitational field, knots may on these grounds be expected quite generally, and one may conjecture that $SU_q(2)$ plays the role of a 
universal hidden symmetry.

I thank A. C. Cadavid and C. Fronsdal for useful comments.

\vskip.5cm

{\bf References.}

\vskip.3cm

{\bf 1.} R. Finkelstein, Lett. Math. Phys. {\bf 62}, 199 (2002).

{\bf 2.} R. Finkelstein, J. Math. Phys. {\bf 37}, 953 (1996).

{\bf 3.} R. Finkelstein, Int. J. Mod. Phys. A{\bf 18}, 627 (2003).

{\bf 4.} S. L. Woronowicz, RIMS, Kyoto {\bf 23}, 112 (1987).

{\bf 5.} L. H. Kauffman, Int. J. Mod. Phys. A{\bf 5}, 93 (1990).

{\bf 6.} G. T. Horowitz, ``Ashtekar's Approach to Quantum Gravity",
in {\it Strings and 
Symmetries 1991}, eds. Berkovits {\it et al.}, World Scientific (1991).

{\bf 7.} C. Rovelli and L. Smolin, Nuc. Phys. B{\bf 331}, 80 (1990).

\end{document}